\documentclass[12pt,epsf]{article}
\usepackage{graphicx}
\usepackage{amssymb,amsmath}

\begin{document}

\def\a{\alpha}
\def\b{\beta}
\def\c{\varepsilon}
\def\d{\delta}
\def\e{\epsilon}
\def\f{\phi}
\def\g{\gamma}
\def\h{\theta}
\def\k{\kappa}
\def\l{\lambda}
\def\m{\mu}
\def\n{\nu}
\def\p{\psi}
\def\q{\partial}
\def\r{\rho}
\def\s{\sigma}
\def\t{\tau}
\def\u{\upsilon}
\def\v{\varphi}
\def\w{\omega}
\def\x{\xi}
\def\y{\eta}
\def\z{\zeta}
\def\D{\Delta}
\def\G{\Gamma}
\def\H{\Theta}
\def\L{\Lambda}
\def\F{\Phi}
\def\P{\Psi}
\def\S{\Sigma}
\def\BR{{\rm Br}}
\def\o{\over}
\def\beq{\begin{eqnarray}}
\def\eeq{\end{eqnarray}}
\newcommand{\nn}{\nonumber \\}
\newcommand{\gsim}{ \mathop{}_{\textstyle \sim}^{\textstyle >} }
\newcommand{\lsim}{ \mathop{}_{\textstyle \sim}^{\textstyle <} }
\newcommand{\vev}[1]{ \left\langle {#1} \right\rangle }
\newcommand{\bra}[1]{ \langle {#1} | }
\newcommand{\ket}[1]{ | {#1} \rangle }
\newcommand{\EV}{ {\rm eV} }
\newcommand{\KEV}{ {\rm keV} }
\newcommand{\MEV}{ {\rm MeV} }
\newcommand{\GEV}{ {\rm GeV} }
\newcommand{\TEV}{ {\rm TeV} }
\def\diag{\mathop{\rm diag}\nolimits}
\def\Spin{\mathop{\rm Spin}}
\def\SO{\mathop{\rm SO}}
\def\O{\mathop{\rm O}}
\def\SU{\mathop{\rm SU}}
\def\U{\mathop{\rm U}}
\def\Sp{\mathop{\rm Sp}}
\def\SL{\mathop{\rm SL}}
\def\tr{\mathop{\rm tr}}

\newcommand{\bear}{\begin{array}}  
\newcommand {\eear}{\end{array}}
\newcommand{\la}{\left\langle}  
\newcommand{\ra}{\right\rangle}
\newcommand{\non}{\nonumber}  
\newcommand{\ds}{\displaystyle}
\newcommand{\red}{\textcolor{red}}
\def\ubl{U(1)$_{\rm B-L}$}
\def\REF#1{(\ref{#1})}
\def\lrf#1#2{ \left(\frac{#1}{#2}\right)}
\def\lrfp#1#2#3{ \left(\frac{#1}{#2} \right)^{#3}}
\def\OG#1{ {\cal O}(#1){\rm\,GeV}}

\def\TODO#1{ {\bf ($\clubsuit$ #1 $\clubsuit$)} }


\baselineskip 0.7cm

\begin{titlepage}

\begin{flushright}
IPMU-13-0010
\end{flushright}

\vskip 1.35cm
\begin{center} 
{\large \bf 
Focus Point in Gaugino Mediation
} \\
\vspace{2.5mm}
{\large \sl $-$ Reconsideration of the Fine-tuning Problem $-$}
\vskip 1.2cm

{Tsutomu T. Yanagida and Norimi Yokozaki}

\vskip 0.4cm

{\it
Kavli Institute for the Physics and Mathematics of the Universe,\\
Todai Institutes for Advanced Study, University of Tokyo,\\
Kashiwa 277-8583, Japan\\
}

\vskip 1.5cm

\abstract{
We reconsider the fine-tuning problem in SUSY models, motivated by the recent observation of the relatively  heavy Higgs boson and non-observation of the SUSY particles at the LHC. Based on this thought, we demonstrate a focus point-like behavior in a gaugino mediation model, and show that the fine-tuning is indeed reduced to about 2$\%$ level if the ratio of the gluino mass to wino mass is about 0.4 at the GUT scale. We show that such a mass ratio may  arise naturally in a product group unification model without the doublet-triplet splitting problem. This fact suggests that the fine-tuning problem crucially depends on the physics at the high energy scale.
 }
\end{center}
\end{titlepage}

\setcounter{page}{2}

\section{Introduction}

The Higgs boson mass is a good probe of the supersymmetry (SUSY) breaking scale in the minimal SUSY standard model (MSSM) \cite{OYY}.
The observed Higgs boson mass of around 125 GeV \cite{ATLAS, CMS} suggests, together with non-discovery of SUSY particles at the LHC, that the SUSY breaking scale is considerably higher than the electroweak scale. This already raises doubt of the low scale SUSY as a solution to the hierarchy problem. In fact, we need a fine-tuning at the level of $0.1-0.01 \%$ to reproduce the correct electroweak symmetry breaking scale if the squark and gluino masses are of order a few TeV.

The purpose of this paper is to argue that the issue of fine-tuning crucially depends on physics at a high energy, say GUT scale. A famous example is so called ``Focus Point SUSY"~\cite{FMM1} (see also \cite{focus_recent0, focus_recent} for recent discussions) in gravity mediation models. In this scenario, small gaugino masses and certain relations among stop masses, the up-type Higgs soft mass and the trilinear coupling of the stop, $A_t$, are assumed. As a result, the Higgs boson mass of around $125$ GeV can be accommodated within about $1\%$ tuning. Although the essential point of ``Focus Point SUSY" is attractive, the relations among the scalar squared masses and $A_t^2$ seem not so simple; the Kahler potential should be carefully chosen in order to reduce the fine-tuning to $1\%$ level.

In this paper, we point out that the focus point like behavior also occurs in gaugino mediation models~\cite{IKYY, gaugino} with one simple relation; the required fine-tuning is indeed reduced significantly, depending on a gaugino mass ratio $M_3/M_2$ at the GUT scale. Here, $M_3$ and $M_2$ are masses of gluino and wino at the GUT scale, respectively.
It may be interesting that the mass ratio could be a parameter independent of SUSY breaking scale.
We stress that the unnatural looking SUSY is a consequence of physics at high energy scale.

\section{Focus point in gaugino mediation}

The recent analyses \cite{NTY} of the adiabatic solution \cite{Linde} to the Polonyi problem~\cite{Polonyi} in gravity mediation scenario would suggest a small gravitino mass, $m_{3/2}$, compared with the gaugino masses $M_{1/2}$, that is, $m_{3/2} \ll M_{1/2}$, and hence the gaugino mediation model \cite{MYY} is very attractive. Furthermore, it is well known that
the flavor changing neutral current (FCNC) problem is ameliorated substantially in the gaugino mediation models~\cite{IKYY, gaugino}. Motivated by those facts, we consider a gaugino mediation model throughout this paper and point out that the focus point-like behavior occurs with a suitable choice of the ratio of $M_3$ and $M_2$; if the ratio of $M_3/M_2\sim 0.4$, the required fine-tuning can be reduced.\footnote{The reduction of the fine-tuning by adopting non-universal gaugino masses is discussed based on the different assumptions \cite{different}.} 
Note that the bino mass $M_1$ is not important, as shown later.

In our setup, among superfields in MSSM, only gauge kinetic functions have enhanced couplings to the Polonyi field which has a SUSY breaking F-term, and hence the scalar masses, the Higgs B-term and scalar trilinear couplings are much smaller than gaugino masses at the high energy scale~\cite{MYY}. The gravitino is the lightest SUSY particle (LSP) and candidate for a dark matter (see \cite{MYY} for details). 
Let us parameterize the gaugino mediation model as
\begin{eqnarray}
(M_1, M_2, M_3)=M_{1/2} ( r_1, 1, r_3),  \ \ \mu_0,
\end{eqnarray}
where $M_1$, $M_2$ and $M_3$ are the bino, wino and gluino mass at the GUT scale, respectively, and $\mu_0$ denotes the Higgsino mass parameter at the GUT scale. 
Here, the scalar masses, the Higgs B-term as well as the scalar trilinear couplings are neglected for simplicity, and they are induced by renormalization group (RG) evolutions between the GUT scale and the SUSY scale.  The universal gaugino mass corresponds to $r_1=r_3=1$. Here and hereafter, we take $r_1, r_3>0$.

The successful electroweak symmetry breaking occurs with a particular balance among the soft SUSY breaking mass of up- and down-type Higgs ($H_u$ and $H_d$), the Higgs B-term and the SUSY invariant mass $\mu$. Including radiative corrections to the Higgs potential, the electroweak symmetry breaking scale is determined by the following condition:
\begin{eqnarray}
\frac{m_{\hat{Z}}^2}{2}= \frac{\left( m_{H_d}^2 + \mu^2 + \frac{1}{2 v_d}\frac{\partial \Delta V}{\partial v_d} \right)-\left( m_{H_u}^2 + \mu^2 + \frac{1}{2 v_u}\frac{\partial \Delta V}{\partial v_u} \right)\tan^2\beta }{(\tan^2\beta-1)}, \label{eq:ewsb}
\end{eqnarray}
where $v_u$ and $v_d$ are the vacuum expectation values of $H_u^0$ and $H_d^0$, respectively, and $\Delta V$ is the radiative correction to the Higgs potential. 
The soft mass squared of $H_u$ and $H_d$ are denoted by $m_{H_u}^2$ and $m_{H_d}^2$, respectively, and $\mu$ is the Higgsino mass parameter at the SUSY scale. The electroweak symmetry breaking scale is, in principle, determined by Eq.\,(\ref{eq:ewsb}) although it is fixed to reproduce  $m_{\hat{Z}} \simeq 91.2$ GeV~\cite{PDG}.
Neglecting $\Delta V$ and the terms suppressed by $\tan^2 \beta$, Eq.\,(\ref{eq:ewsb}) is simplified to $m_{\hat{Z}}^2 \sim -2 m_{H_u}^2 - 2 \mu^2$. This clearly shows the dependence of $m_{\hat{Z}}^2$ on $m_{H_u}^2$ and $\mu^2$.

\begin{figure}[t]
\begin{center}
\includegraphics[width=11cm]{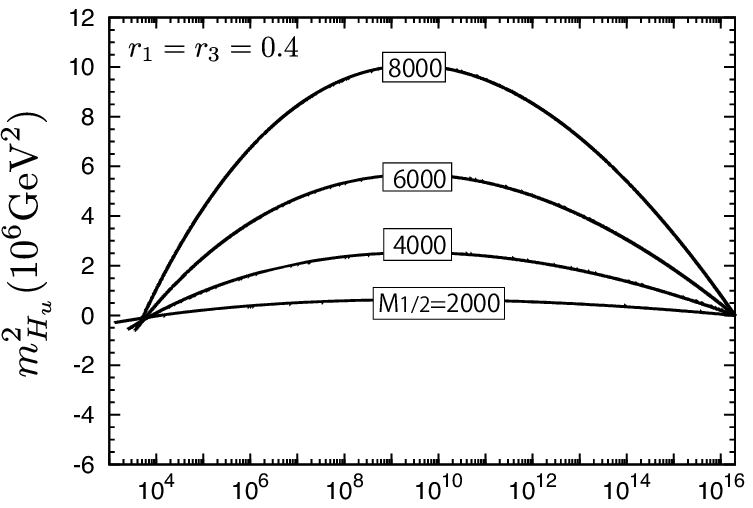}
\includegraphics[width=11cm]{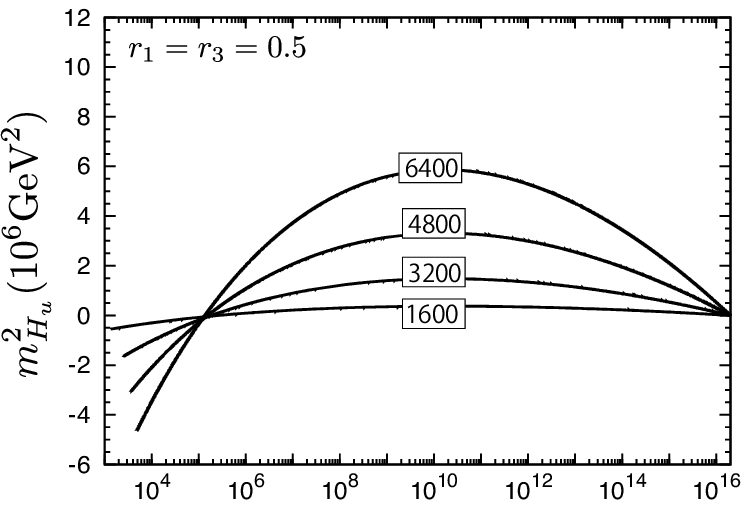}
\caption{$m_{H_u}^2$ as a function of the renormalization scale (${\rm GeV}$). The ratios $r_1$ and $r_3$ are taken as $r_1=r_3=0.4$ ($r_1=r_3=0.5$) on the upper (lower) panel. The four solid lines correspond to $M_{1/2}=8000$, 4000, 6000, 2000\GEV from top to bottom on the upper panel, while $M_{1/2}=6400$, 4800, 3200, 1600\GEV on the lower panel. Here, $\tan\beta=20$, $\alpha_S(m_Z)=0.1184$ and $m_t({\rm pole})=173.2\,{\rm GeV}$. 
}
\label{fig:focus}
\end{center}
\end{figure}

Since there are only three input parameters for the SUSY breaking, $M_1$, $M_2$ and $M_3$, all soft SUSY breaking parameters including $m_{H_u}^2$ can be written as a function of $M_1$, $M_2$ and $M_3$. For instance, taking $\tan\beta=20$, $m_{H_u}^2$ at $3$\,TeV (the renormalization scale) is approximately given by
\begin{eqnarray}
m_{H_u}^2(3\, {\rm TeV}) &\simeq& -1.21 M_3^2 + 0.21 M_2^2  - 0.017 M_1 M_3  -0.10 M_2 M_3 \nonumber \\ && + 0.009 M_1^2  -0.006 M_1 M_2, 
\end{eqnarray}
where the two-loop renormalization group equations~\cite{2loopRGE} are used. We obtain $m_{H_u}^2 \simeq -0.006 M_{1/2}^2$ for $r_1=r_3=0.38$, while $m_{H_u}^2 \simeq -1.12 M_{1/2}^2$ for $r_1=r_3=1.0$.
This indicates that the fine-tuning can be reduced with a certain choice of $r_3$, that is, the ratio of $M_3$ to $M_2$. Notice that the coefficients of the terms proportional to $M_1$ are small in  most of the viable region,\footnote{The stau becomes tachyonic for $r_1 \ll r_3$ unless $\tan\beta$ is small.} and hence, their contributions to $m_{H_u}^2$ are not important as long as $M_1 \sim M_3$.
In Fig.~\ref{fig:focus}, we show the focus point-like behavior for different choice of $r_3$ (and $r_1$). The scale where $m_{H_u}^2$ vanish is shifted to the low-scale as $r_3$ becomes small, and hence, by taking smaller value of $r_3$, it is expected that the amount of the fine-tuning is reduced.

In order to evaluate the degrees of fine-tuning, we adapt the following fine-tuning measure:~\footnote{
The definition of the fine-tuning measure Eq.(\ref{eq:ft}) differs by a factor of 2, compared to the original definition~\cite{FTM}. This definition may be more natural, considering $\Delta_\mu \sim 2 \mu^2/(91.2\, {\rm GeV})^2$ (see also \cite{focus_recent} for comments on $\Delta$). }
\begin{eqnarray}
\Delta \equiv{\rm max}\{\Delta_a\}, \ \ \Delta_a \equiv \left|  \frac{\partial \ln m_{\hat{Z}}^2}{\partial \ln a^2}\right|, \label{eq:ft}
\end{eqnarray}
where $a$ is a parameter at the GUT scale and $a=M_{1/2}$ and $\mu_0$ in our model. Notice that $\Delta_\mu$ is always $ \sim 2 \mu^2/(91.2\, {\rm GeV})^2$, since the SUSY mass parameter $\mu$ is almost unchanged during the RG evolution between the GUT scale to the SUSY scale, i.e., $\mu \simeq \mu_0$, and hence a small $\Delta_\mu$ simply means a small $\mu$. On the other hand, roughly speaking, a small $\Delta_{M_{1/2}}$ means a small change of $m_{H_u}^2$, and hence, a small $\mu$ does not always correspond to a small fine-tuning.

\begin{figure}[t]
\begin{center}
\includegraphics[width=11cm]{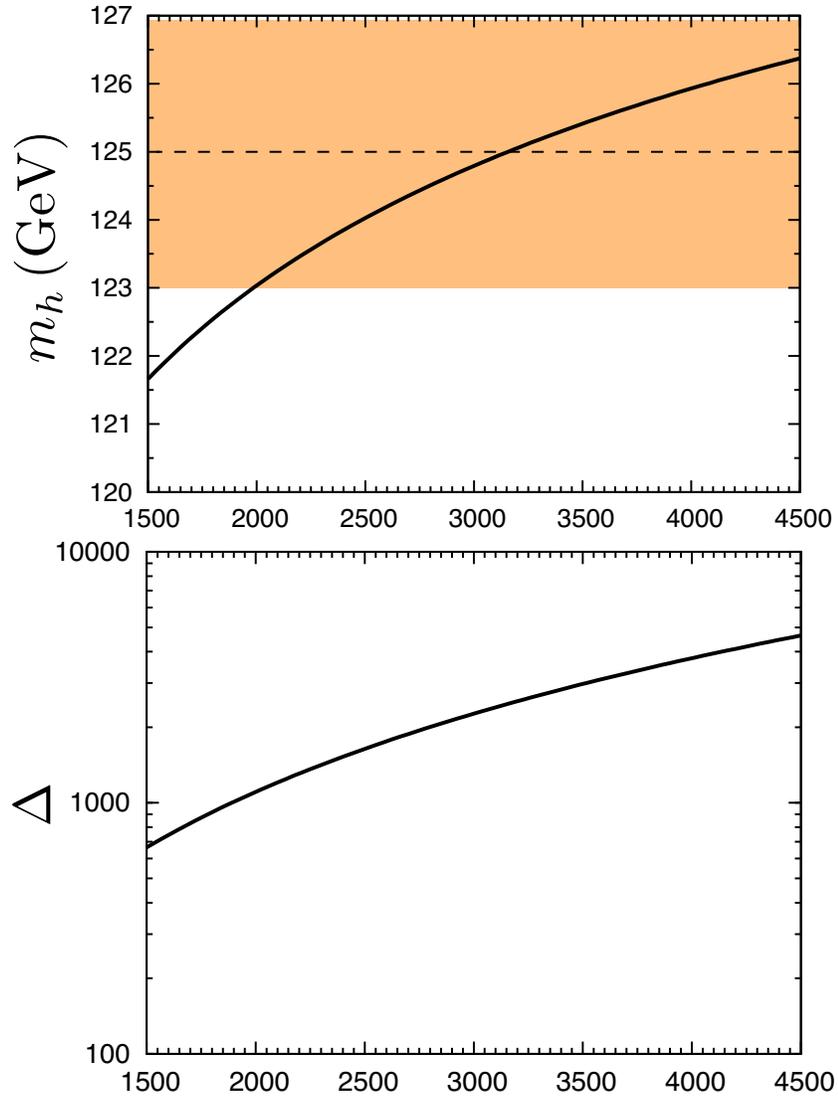}
\caption{The Higgs boson mass and $\Delta$ as a function of $M_{1/2}$ in the case of the universal gaugino mass. The other parameters, $\tan\beta$, $\alpha_S(m_Z)$ and $m_t({\rm pole})$ are same as in Fig.~{\ref{fig:focus}}. 
}
\label{fig:univ}
\end{center}
\end{figure}

First, we show results of the universal gaugino mass case, i.e., $r_1=r_3=1.0$ in Fig.~\ref{fig:univ}. The Higgs boson mass, $m_h$, is shown on the upper panel, while $\Delta$ is shown on the lower panel. The Higgs boson mass  and the mass spectrum of the SUSY particles are calculated by {\tt SuSpect} package~\cite{suspect}. 
The Higgs boson mass of $123\  (125)$ GeV is obtained with $M_{1/2} \simeq 2000\  (3100)$ GeV.\footnote{We have checked that $M_{1/2}=2000$ GeV can be consistent with the Higgs boson mass of 125 GeV using {\tt FeynHiggs} package~\cite{feynhiggs}.} The corresponding fine-tuning parameter $\Delta$ is  $\simeq 1090\, (2520)$, that is, $0.09\, (0.04)\%$ tuning. The stop mass, $m_{\tilde{t}} \equiv (m_{\tilde{t}_1} + m_{\tilde{t}_2})/2$, is predicted as $m_{\tilde{t}} \simeq 3250 \, (4890)$ GeV. Here, $m_{\tilde{t}_1}$ and $m_{\tilde{t}_2}$ are the light and heavy stop mass, respectively. Considering $2-3$ GeV uncertainty of the Higgs boson mass calculation, we need at least $0.1\%$ fine-tuning.

\begin{figure}[t]
\begin{center}
\includegraphics[width=12cm]{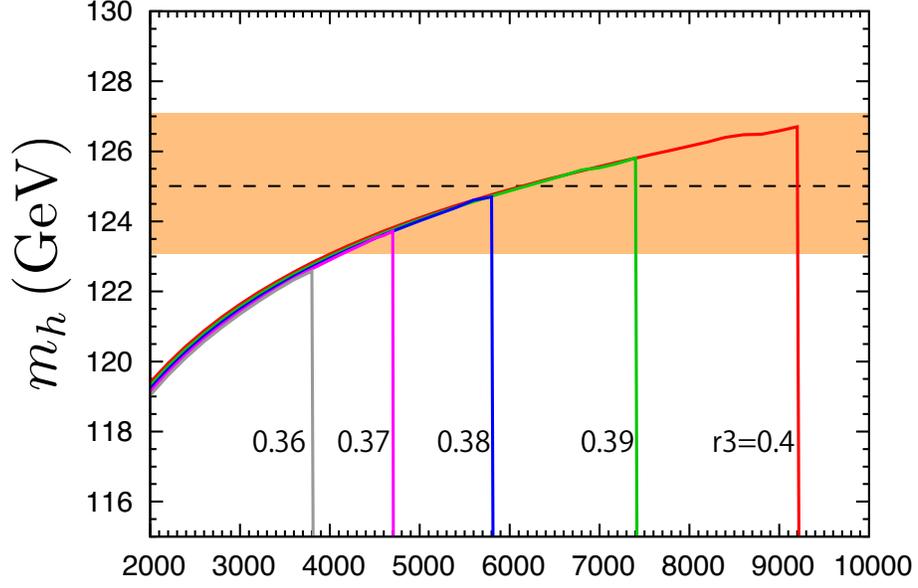}
\caption{The Higgs boson mass as a function of $M_{1/2}$ for different $r_3$. The sudden drop of $m_h$  corresponds to unsuccessful EWSB.}
\label{fig:higgs}
\end{center}
\end{figure}

\begin{figure}[t]
\begin{center}
\includegraphics[width=12cm]{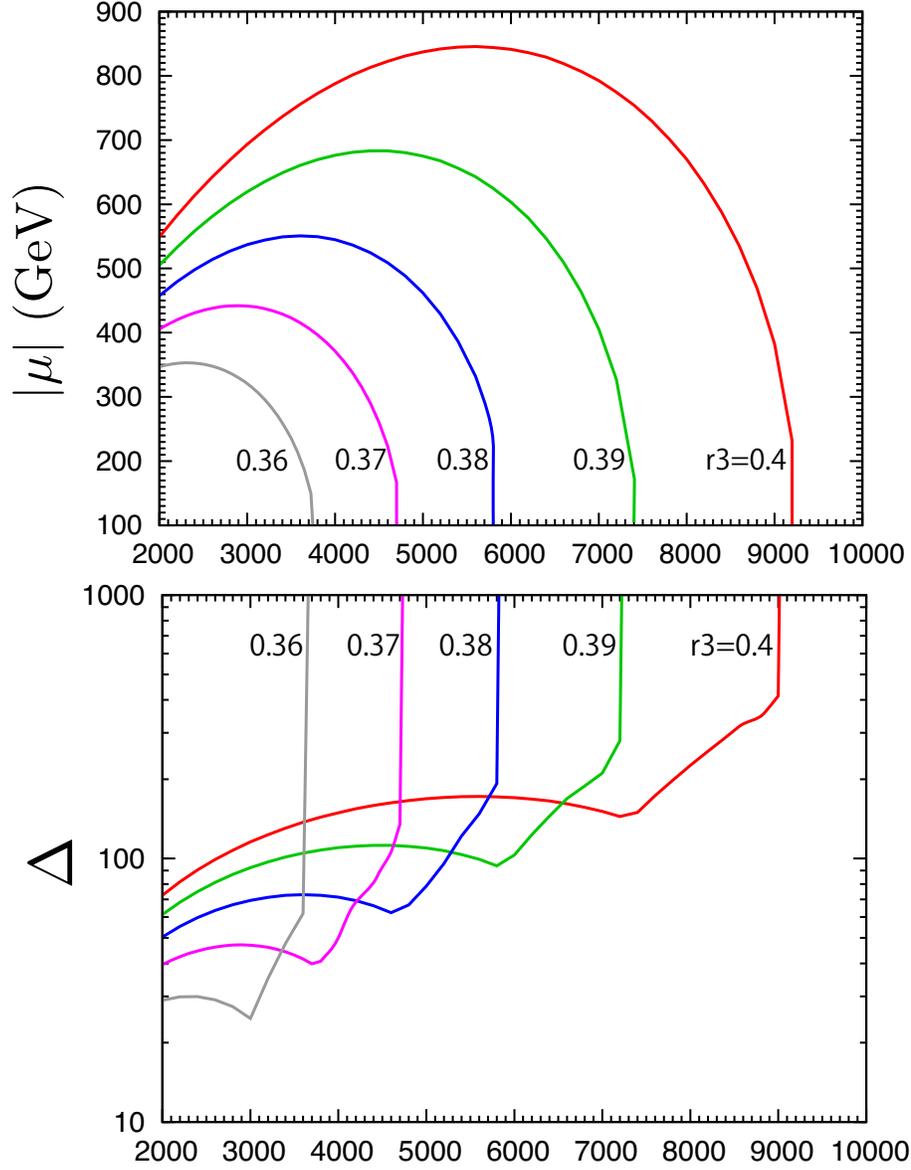}
\caption{$|\mu|$ and $\Delta$ as a function of $M_{1/2}$ for different $r_3$. The ratio $r_1$ is taken as $r_1=0.4$. Other parameters are same as in Fig.~\ref{fig:focus}. The vertical rise of $\Delta$ corresponds to  unsuccessful EWSB.}
\label{fig:mu}
\end{center}
\end{figure}

In the case of non-universal gaugino masses, the fine-tuning is reduced significantly due to the focus point-like behavior. In Fig.~{\ref{fig:higgs}}, the Higgs boson mass as a function of $M_{1/2}$ is shown for different $r_3$. The ratio $r_1$ is taken as $r_1=0.4$. The slight change of the ratio $r_3$ does not affect the Higgs boson mass significantly. The calculated Higgs boson mass is 123 (125) GeV for $M_{1/2} \simeq 4100$ (6200) GeV. The Higgsino mass $\mu$ and $\Delta$ are shown in Fig.~\ref{fig:mu}. Sharp bends of $\Delta$ in the lower panel (e.g., $r=0.36$ and $M_{1/2} \simeq 3100$\,GeV) correspond to the change of the dominant contributions to $\Delta$. In the region with small $M_{1/2}$, $\Delta$ is simply determined by the size of $\mu$ parameter. As $M_{1/2}$ gradually increases, $|\mu|$ becomes small. However, $(\partial \ln m_{\hat Z}^2)/(\partial \ln M_{1/2}^2)$ dominates $\Delta$, and the fine-tuning becomes worse. This change is also reflected in the steep slope of $|\mu|$; the small $|\mu|$ is necessary for small $\Delta$ but it is not sufficient. It is noticed that the fine-tuning measure is reduced to $\Delta \simeq 60$ (123) for $r_3=0.37$ (0.39), where the gaugino mass is taken as $M_{1/2} \simeq 4100$ (6200) GeV. The observed Higgs boson mass of around 125 GeV can be consistent with about $2\, \%$ tuning. The detailed mass spectra are shown in Table.~{\ref{table:mass}}. Since  some of the squark masses can be smaller than 3 TeV, they may be observed at LHC with $\sqrt{s}=14$\,TeV. In addition, the lightest stau, chargino and neutralino can be around 350 GeV, which may be target of future linear collider experiments.

\begin{table}[t!]
  \begin{center}
    \begin{tabular}{  c | c  }
    $M_{1/2}$ & 4100 GeV \\
    $r_1, r_3$ & 0.4, 0.37 \\
    $\tan\beta$ & 20 \\
    \hline
\hline    
    $\mu$ & $-355$ GeV \\
    $\Delta$ & 60 \\
\hline
    $m_{h}$ & 123 GeV \\
    $m_{\rm gluino}$ & 3280 GeV \\
    $m_{\tilde{t}_1}$ & 1760 GeV \\
    $m_{\tilde{t}_2}$ & 3420 GeV \\
    $ A_t $ & $-3100$ GeV \\
    $m_{\tilde{q}}  $ & 2770-3750 GeV \\
    $m_{\tilde{e}_L} (m_{\tilde{\mu}_L})$ & 2610 GeV\\
    $m_{\tilde{e}_R} (m_{\tilde{\mu}_R})$ & 600 GeV \\
    $m_{\tilde{\tau}_1}$ & 375 GeV \\
    $m_{\chi_1^0}$ & 361 GeV \\
     $m_{\chi_1^{\pm}}$ & 364 GeV \\
    \end{tabular}
        \begin{tabular}{  c | c  }
    $M_{1/2}$ & 6200 GeV \\
    $r_1, r_3$ & 0.4, 0.39 \\
    $\tan\beta$ & 20 \\
    \hline
\hline    
    $\mu$ & $-578$ GeV \\
    $\Delta$ & 123 \\
\hline
    $m_{h}$ & 125 GeV \\
    $m_{\rm gluino}$ & 5050 GeV \\
    $m_{\tilde{t}_1}$ & 2790 GeV \\
    $m_{\tilde{t}_2}$ & 5190 GeV \\
    $ A_t $ & $-4700$ GeV \\
    $m_{\tilde{q}}  $ & 4240-5670 GeV \\
    $m_{\tilde{e}_L} (m_{\tilde{\mu}_L})$ & 3890 GeV\\
    $m_{\tilde{e}_R} (m_{\tilde{\mu}_R})$ & 899 GeV \\
    $m_{\tilde{\tau}_1}$ & 577 GeV \\
    $m_{\chi_1^0}$ & 594 GeV \\
     $m_{\chi_1^{\pm}}$ & 596 GeV \\
    \end{tabular}
    \caption{The mass spectrum and $\Delta$. The scalar trilinear coupling of the stop is denoted by $A_t$.
    Here, the gravitino is the LSP.
     }
  \label{table:mass}
  \end{center}
\end{table}

\section{Conclusions and discussion}

In this paper we have shown that the required fine tuning is substantially reduced at the level of $\sim 2\%$ in a gaugino mediation model if the ratio of the gluino mass to the wino mass at the GUT scale is about 0.4.\footnote{If the bino mass is taken to be larger, the fine-tuning becomes further (but slightly) reduced.} The Higgs boson mass of around 125 GeV can be explained without a severe fine-tuning, even if the colored SUSY particles are as heavy as a few TeV. 

The deviation of the universal gaugino mass is clearly inconsistent with the minimal SUSY GUT scenario. However, we show in this section that the required mass ratio, $M_3/M_2\sim 0.4$, is even natural in one of the product group unification (PGU) models \cite{PGU, PGU2}, which were proposed to solve the doublet-triplet splitting problem in the minimal SUSY GUT.

Here, we consider the $SU(5)_{\rm GUT}\times U(2)_H$ PGU model \cite{PGU2}, where $U(2)_H \simeq  SU(2)_H \times U(1)_H$. In this model, $SU(5)_{\rm GUT}\times U(2)_H$ breaks down to the standard model gauge group at the GUT scale without spoiling the gauge coupling unification. 
As a result, the vector superfield of $SU(2)_L$ ($U(1)_Y$) becomes a mixture of those of $SU(2)_{\rm GUT} \subset SU(5)_{\rm GUT}$ ($U(1)_{\rm GUT} \subset SU(5)_{\rm GUT}$) and $SU(2)_H$ ($U(1)_H$), and hence, the gaugino masses become non-universal. The bino, wino and gluino masses at the GUT scale are given by~\footnote{See also \cite{PGU_HCM} for a similar discussion in $SU(5)_{\rm GUT}\times U(3)_H$ PGU model.}
\begin{eqnarray}
M_1 &\simeq& M_{\rm GUT} + (3/5) g_{\rm GUT}^2 M_{H1} / {g_{H1}^2}, \nonumber \\
M_2 &\simeq& M_{\rm GUT} + g_{\rm GUT}^2 M_{H2} / {g_{H2}^2}, \nonumber \\
M_3 &\simeq& M_{\rm GUT},
\end{eqnarray}
where $M_{\rm GUT}$, $M_{H1}$ and $M_{H2}$ ($g_{\rm GUT}$, $g_{H1}$ and $g_{H2}$) are gaugino masses (gauge couplings) of $SU(5)_{\rm GUT}$, $U(1)_H$ and $SU(2)_H$, respectively.~\footnote{
Here, we take a normalization of $U(1)_H$ such that ${\rm Tr}(t^a t^b)=(1/2)\delta^{ab}$ for the fundamental representation of $U(2)_H$, where $t^0=(1/2){\bf 1}_{2 \times 2}$ and $t^{1,2,3}$ are generators of $SU(2)_H$~\cite{watari_ibe}.} 
Therefore, if $M_{H2}/g_{H2}^2$ is comparable to $M_{\rm GUT}/g_{\rm GUT}^2$, the desired ratio, $M_3/M_2 \sim 0.4$ can be obtained. The focus point in gaugino mediation discussed in this paper may be naturally explained in more fundamental physics of the PGU model at the GUT scale.

Finally, let us comment on the constraint from the electric dipole moment (EDM) of the electron.\footnote{
The EDM of the neutron gives an similar constraint, which can be also avoided.
} In the PGU model, the phases of the gaugino masses are not aligned in general, and potentially dangerous CP violating phases are generated. The SUSY contributions to the EDM are approximately proportional to the following combinations of the CP violating phases:
\begin{eqnarray}
\theta_i = {\rm Arg} (\mu (B \mu)^* \tilde{M}_i) \simeq {\rm Arg} (\mu (B \mu)^* {M}_i), \label{eq:theta}
\end{eqnarray}
where $\tilde{M}_i$ is the gaugino mass at the SUSY scale. The Higgs B parameter at the SUSY scale is approximately given by
\begin{eqnarray}
B(3\, {\rm TeV}) \simeq -0.017 M_1 -0.300 M_2 + 0.290 M_3,
\end{eqnarray}
for $\tan\beta=20$. As a reference, we take the phases of the gaugino masses as ${\rm Arg} (M_1) = {\rm Arg} (M_3) = 0.1$ and ${\rm Arg} (M_2)=0$. Consequently, the generated CP violating phases (\ref{eq:theta}) are $\theta_{1,3} \simeq 0.15-\pi$ and $\theta_2 \simeq 0.05-\pi$, and the predicted electron EDM is $|d_e| \simeq 7.6 \times 10^{-28} \, e\, {\rm cm}$ ($|d_e| \simeq 3.1 \times 10^{-28} \, e\, {\rm cm}$ ) for $M_{1/2} = 4100\, {\rm GeV}, r_1=1.5$ and $r_3=0.37$ ($M_{1/2} = 6200\, {\rm GeV}, r_1=1.5$ and $r_3=0.39$), which is below the current experimental bound, $d_e \lesssim 10^{-27}\, e\, {\rm cm}$~\cite{PDG}.  As we have stated, the change of the bino mass does not affect the focus point-like behavior significantly, that is, the bino can be heavy without an increase of  the fine-tuning. Therefore, the constraint from the EDM can be avoided relatively easily, but still the electron EDM is expected to be seen at feature experiments.

\section*{Acknowledgements}
The work of NY is supported in part by JSPS Research Fellowships for Young Scientists.
This work is also supported by the World Premier International Research Center Initiative (WPI Initiative), MEXT, Japan.



\begin{thebibliography}{99}

\bibitem{OYY} 
  Y.~Okada, M.~Yamaguchi and T.~Yanagida,
  Prog.\ Theor.\ Phys.\  {\bf 85}, 1 (1991);
   Y.~Okada, M.~Yamaguchi and T.~Yanagida,
  Phys.\ Lett.\ B {\bf 262}, 54 (1991);
    J.~R.~Ellis, G.~Ridolfi and F.~Zwirner,
  Phys.\ Lett.\ B {\bf 257}, 83 (1991);
  H.~E.~Haber and R.~Hempfling,
  Phys.\ Rev.\ Lett.\  {\bf 66}, 1815 (1991);
  J.~R.~Ellis, G.~Ridolfi and F.~Zwirner,
  Phys.\ Lett.\ B {\bf 262}, 477 (1991).

%
\bibitem{ATLAS} 
 G.~Aad {\it et al.}  [ATLAS Collaboration],
  Phys.\ Lett.\ B {\bf 716}, 1 (2012)
  [arXiv:1207.7214 [hep-ex]].

\bibitem{CMS}
 S.~Chatrchyan {\it et al.}  [CMS Collaboration],
  Phys.\ Lett.\ B {\bf 716}, 30 (2012)
  [arXiv:1207.7235 [hep-ex]].
  
\bibitem{FMM1}
  J.~L.~Feng, K.~T.~Matchev and T.~Moroi,
  Phys.\ Rev.\ Lett.\  {\bf 84}, 2322 (2000)
  [hep-ph/9908309];
  J.~L.~Feng, K.~T.~Matchev and T.~Moroi,
  Phys.\ Rev.\ D {\bf 61}, 075005 (2000)
  [hep-ph/9909334].

  \bibitem{focus_recent0}
  J.~L.~Feng, K.~T.~Matchev and D.~Sanford,
  Phys.\ Rev.\ D {\bf 85}, 075007 (2012)
  [arXiv:1112.3021 [hep-ph]].
  
  \bibitem{focus_recent}
 J.~L.~Feng and D.~Sanford,
  Phys.\ Rev.\ D {\bf 86}, 055015 (2012)
  [arXiv:1205.2372 [hep-ph]].


\bibitem{IKYY}
K.~Inoue, M.~Kawasaki, M.~Yamaguchi and T.~Yanagida,
  Phys.\ Rev.\ D {\bf 45}, 328 (1992).
 \bibitem{gaugino}
  D.~E.~Kaplan, G.~D.~Kribs and M.~Schmaltz,
  Phys.\ Rev.\ D {\bf 62}, 035010 (2000)
  [hep-ph/9911293];
   Z.~Chacko, M.~A.~Luty, A.~E.~Nelson and E.~Ponton,
  JHEP {\bf 0001}, 003 (2000)
  [hep-ph/9911323].
  
\bibitem{NTY} 
  K.~Nakayama, F.~Takahashi and T.~T.~Yanagida,
  Phys.\ Rev.\ D {\bf 84}, 123523 (2011)
  [arXiv:1109.2073 [hep-ph]];
%
  K.~Nakayama, F.~Takahashi and T.~T.~Yanagida,
  Phys.\ Rev.\ D {\bf 86}, 043507 (2012)
  [arXiv:1112.0418 [hep-ph]].


\bibitem{Linde}
  A.~D.~Linde,
  Phys.\ Rev.\ D {\bf 53}, 4129 (1996)
  [hep-th/9601083].
  
 \bibitem{Polonyi}
  G.~D.~Coughlan, W.~Fischler, E.~W.~Kolb, S.~Raby, G.~G.~Ross,
  Phys.\ Lett.\  {\bf B131}, 59 (1983);
  J.~R.~Ellis, D.~V.~Nanopoulos, M.~Quiros,
  Phys.\ Lett.\  {\bf B174}, 176 (1986);
  A.~S.~Goncharov, A.~D.~Linde, M.~I.~Vysotsky,
  Phys.\ Lett.\  {\bf B147}, 279 (1984).


\bibitem{MYY}
  T.~Moroi, T.~T.~Yanagida and N.~Yokozaki,
  arXiv:1211.4676 [hep-ph].
 

\bibitem{different}
 G.~L.~Kane and S.~F.~King,
  Phys.\ Lett.\ B {\bf 451}, 113 (1999)
  [hep-ph/9810374];
H.~Abe, T.~Kobayashi and Y.~Omura,
  Phys.\ Rev.\ D {\bf 76}, 015002 (2007)
  [hep-ph/0703044 [hep-ph]];
   S.~P.~Martin,
  Phys.\ Rev.\ D {\bf 75}, 115005 (2007)
  [hep-ph/0703097 [hep-ph]];
   D.~Horton and G.~G.~Ross,
  Nucl.\ Phys.\ B {\bf 830}, 221 (2010)
  [arXiv:0908.0857 [hep-ph]];
  J.~E.~Younkin and S.~P.~Martin,
  Phys.\ Rev.\ D {\bf 85}, 055028 (2012)
  [arXiv:1201.2989 [hep-ph]];
  F.~Brummer and W.~Buchmuller,
  JHEP {\bf 1205}, 006 (2012)
  [arXiv:1201.4338 [hep-ph]];
  I.~Gogoladze, F.~Nasir and Q.~Shafi,
  arXiv:1212.2593 [hep-ph].
\bibitem{PDG}
J. Beringer et al. (Particle Data Group), Phys. Rev. D86, 010001 (2012)

\bibitem{2loopRGE} 
S.P.~Martin and M.T.~Vaughn,
  Phys.\ Rev.\ D {\bf 50}, 2282 (1994)
  [hep-ph/9311340].
  
  \bibitem{FTM}
J.~R.~Ellis, K.~Enqvist, D.~V.~Nanopoulos and F.~Zwirner,
  Mod.\ Phys.\ Lett.\ A {\bf 1}, 57 (1986);
   R.~Barbieri and G.~F.~Giudice,
  Nucl.\ Phys.\ B {\bf 306}, 63 (1988).


\bibitem{suspect}
    A.~Djouadi, J.~-L.~Kneur and G.~Moultaka,
  Comput.\ Phys.\ Commun.\  {\bf 176}, 426 (2007)
  [hep-ph/0211331].

\bibitem{feynhiggs}
    S.~Heinemeyer, W.~Hollik and G.~Weiglein,
  Comput.\ Phys.\ Commun.\ \ {\bf 124}, 76  (2000)
  [hep-ph/9812320].
  S.~Heinemeyer, W.~Hollik and G.~Weiglein,
  Eur.\ Phys.\ J.\ C\ {\bf 9}, 343  (1999)
  [hep-ph/9812472].
  G.~Degrassi, S.~Heinemeyer, W.~Hollik, P.~Slavich and G.~Weiglein,
  Eur.\ Phys.\ J.\ C\ {\bf 28}, 133  (2003)
  [hep-ph/0212020].
  M.~Frank, T.~Hahn, S.~Heinemeyer, W.~Hollik, H.~Rzehak and G.~Weiglein,
  JHEP\ {\bf 0702}, 047  (2007)
  [hep-ph/0611326].
  
\bibitem{PGU}

  T.~Yanagida,
  Phys.\ Lett.\ B {\bf 344}, 211 (1995)
  [hep-ph/9409329];
    T.~Hotta, K.~I.~Izawa and T.~Yanagida,
  Phys.\ Rev.\ D {\bf 53}, 3913 (1996)
  [hep-ph/9509201];
  T.~Hotta, K.~I.~Izawa and T.~Yanagida,
  Phys.\ Rev.\ D {\bf 54}, 6970 (1996)
  [hep-ph/9602439];
   J.~Hisano and T.~Yanagida,
  Mod.\ Phys.\ Lett.\ A {\bf 10}, 3097 (1995)
  [hep-ph/9510277];
    K.~I.~Izawa and T.~Yanagida,
  Prog.\ Theor.\ Phys.\  {\bf 97}, 913 (1997)
  [hep-ph/9703350].

\bibitem{PGU2}
  T.~Watari and T.~Yanagida,
  hep-ph/0208107;
   T.~Watari and T.~Yanagida,
  Phys.\ Rev.\ D {\bf 70}, 036009 (2004)
  [hep-ph/0402160].
 
\bibitem{PGU_HCM}
N.~Arkani-Hamed, H.~-C.~Cheng and T.~Moroi,
  Phys.\ Lett.\ B {\bf 387}, 529 (1996)
  [hep-ph/9607463].
  
  \bibitem{watari_ibe}
  See, e.g.,
   M.~Ibe and T.~Watari,
  Phys.\ Rev.\ D {\bf 67}, 114021 (2003)
  [hep-ph/0303123].
  
  
\end{thebibliography}
\end{document}